# Direct measurement of vorticity using tracer particles with internal markers

Jiaqi Li[1,2], Lei Feng[1], Chinmayee Panigrahi[2], Jiarong Hong[1,2, *]

[1]Saint Anthony Falls Laboratory, University of Minnesota, Minneapolis, MN, 55455, USA
[2]Department of Mechanical Engineering, University of Minnesota, Minneapolis, MN, 55455, USA

* Email address of the corresponding author: jhong@umn.edu

**Abstract**

Current experiment techniques for vorticity measurement suffer from limited spatial and temporal resolution to resolve the small-scale eddy dynamics in turbulence. In this study, we develop a new method for direct vorticity measurement in fluid flows based on digital inline holography (DIH). The DIH system utilizes a collimated laser beam to illuminate the tracers with internal markers and a digital sensor to record the generated holograms. The tracers made of the polydimethylsiloxane (PDMS) prepolymer mixed with internal markers are fabricated using a standard microfluidic droplet generator. A rotation measurement algorithm is developed based on the 3D location reconstruction and tracking of the internal markers and is assessed through synthetic holograms to identify the optimal parameter settings and measurement range (e.g., rotation rate from 0.3 to 0.7 rad/frame under numerical aperture of imaging of 0.25). Our proposed method based on DIH is evaluated by a calibration experiment of single tracer rotation, which yields the same optimal measurement range. Using von Kármán swirling flow setup, we further demonstrate the capability of the approach to simultaneously measure the Lagrangian rotation and translation of multiple tracers. Our method can measure vorticity in a small region on the order of 100 µm or less and can be potentially used to quantify the Kolmogorov-scale vorticity field in turbulent flows.

## 1. Introduction

Investigating the spatial distribution and temporal evolution of the vorticity field in turbulence is critical for understanding turbulence beyond statistical averaging and stochastic dynamics. It provides a direct way to advance the phenomenological interpretation of turbulence as chains of cascading eddies, owing to the inherent nature of vorticity in manifesting the smallest "whirls" in the flow (Jiménez et al. 1993). However, the direct measurement of the vorticity field at small scales in turbulence is a formidable challenge. Therefore, as of today, investigations of this kind are almost exclusively conducted using direct numerical simulation (DNS) (e.g., Rogers & Moin 1987; She et al. 1990; Sreenivasan & Antonia 1997; Yang et al. 2010; Lozano-Durán & Jiménez 2014; Cardesa et al. 2017; Elsas et al. 2018).

Although the DNS provides "nearly perfect" data for the theoretical study of turbulence, significant challenges still exist in the data analysis methodology, such as tracking time-evolving geometric structures at various scales in turbulence, characterizing the interaction among these structures, and establishing its linkage with statistical relations (Pradeep & Hussain 2010; Yang et al. 2010). In addition to the limited range of Reynolds numbers, DNS has some other inherent limitations, including ambiguities on proper boundary conditions and sufficient grid size, which could lead to a substantial difference in comparison with high-resolution experimental measurements under specific settings. Specifically, the experiment conducted by Sheng et al. (2009) in the study of near-wall turbulence in a smooth-wall channel flow reported an abundance of pairs of streamwise vortex structures in the buffer layer, inconsistent with the conclusions of several studies using DNS data (e.g., Jiménez 1999; Schoppa & Hussain 2000 & 2002). Furthermore, their study raised concerns about the sufficiency of the DNS resolution in the streamwise direction for resolving the abrupt lifting of small-scale vortex filaments during early stages due to the lack of knowledge on the small-scale structures in turbulence. Therefore, experiments still serve as an indispensable component for validating and revealing missing physics in the simulation. In particular, the tracker-based whole field measurements can provide a natural and powerful solution for investigating time-evolving structures in turbulence. Nowadays, state-of-the-art experimental studies of the vortex structures in turbulence have been conducted using the 2D and 3D particle image/tracking velocimetry (PIV/PTV) (e.g., Guala et al. 2005; Guala et al. 2006; Elsinga & Marusic 2010), which allows us to assess all the relevant terms in Cauchy-Green tensor to probe into the vorticity-strain interaction. However, with such an approach, vorticity is only derived from the gradient of the velocity field. The highest achievable resolution is still two orders lower than what is needed to capture the smallest eddies (e.g., vortex tubes) in the turbulent flow of most laboratory experiments, missing key information to assess the scale interaction during turbulent energy cascade.

Consequently, several new approaches have been proposed to capture vorticity by measuring the rotation of tracers directly. Specifically, Frish & Webb (1981) proposed a direct optical probe for vorticity measurement. Their approach used polydisperse plastic spherical particles (with an average diameter of 21 μm) embedded with one or two hexagonal mirrors made of lead carbonate crystals. They measure the angle and rotation of the mirrors within the particles by tracking the intensity of the reflected light. Following a similar idea, Wu et al. (2018) employed spherical hydrogel tracers of 50 μm in diameter with a planar mirror embedded inside for direct vorticity measurement. As the tracer spins, the mirror inside reflects the laser illumination to generate a bright trajectory on the camera sensor, and the length and radius of curvature of the trajectory are used to determine the rotational rate of each tracer. However, besides the complex tracer fabrication process, such methods only work with a very small number (<10) of tracers in the sample volume and can only obtain snapshots of the tracer rotation. Based on the rotational Doppler effect of the Laguerre-Gaussian beam with orbital angular momentum, Ryabtsev et al. (2016) developed a laser Doppler probe that only measures 1D vorticity at a fixed location in the flow. Using views from multiple cameras and image analysis, a number of studies demonstrated the capability to measure the rotation of spherical and small jack-like particles (Zimmermann et al. 2011a & 2011b; Klein et al. 2012; Marcus et al. 2014). Zimmermann et al. (2011a & 2011b) proposed an approach by tracking the unique textures on the outer surface of a sphere using two cameras from perpendicular views to investigate the rotational intermittency of large (18 mm in diameter) particles in turbulence. Exploring a different path, Klein et al. (2012) employed transparent polymer particles (10 mm in diameter) marked internally with fluorescent tracers to investigate their rotation and translation in fully developed turbulence through particle tracking by a multi-camera system. However, these particles have a size at centimeter-scale, limiting their ability to capture small-scale vortices in turbulence at high Reynolds numbers. Marcus et al. (2014) employed 3D-printed crosses and jacks at a smaller size (around 3 mm) to measure the Lagrangian vorticity and rotation of these particles from four cameras. However, limited by the resolution of

3D printing, the resolution of their vorticimetry is still considerably larger than the Kolmogorov scale of their experiments. To address the gaps between the turbulent theories and available measurement methodology, we present a unique technique based on digital inline holography (DIH) using tracers with internal markers that can directly measure the vorticity field at desired spatial and temporal resolution. The detailed information on the technique is introduced in Section 2, and we assess the proposed methodology in Section 3 with conclusions and discussion in Section 4.

## 2. Experimental Methodology

The setup and the procedure for the proposed measurement technique are shown in Fig. 1, which comprises a digital inline holography (DIH) system to image small spherical tracers with internal markers in a three-dimensional (3D) volume and the corresponding data analysis method to extract the 3D rotation of individual tracers for vorticity measurements. The detailed information about the experimental setup, including tracer fabrication and data analysis, is described in Sections 2.1 and 2.2, respectively.

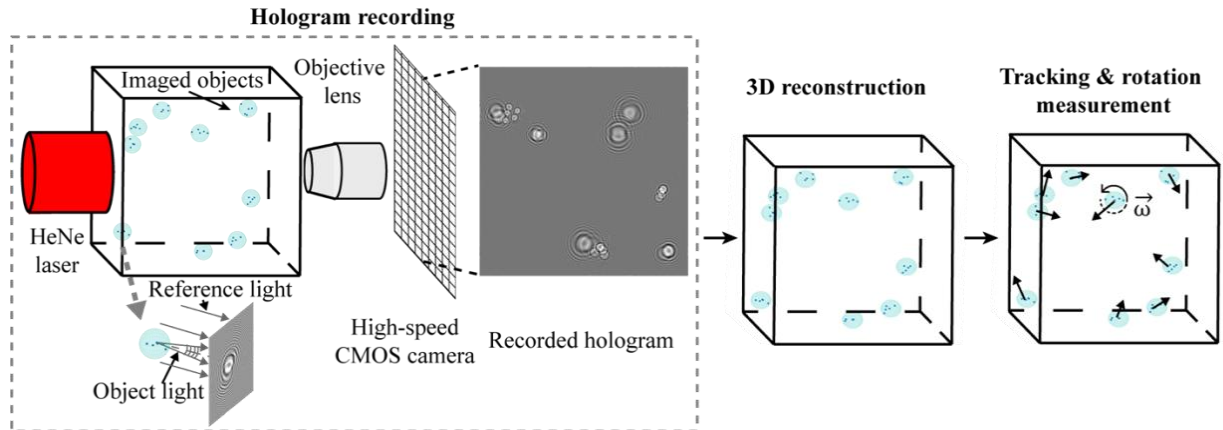

Figure 1: Illustration of the steps for vorticity field measurement based on digital inline holography. The digital inline holography setup is employed for hologram recording (inside the box with dashed lines). When the laser illuminates the 3D sample volume, the scattered light from each marker (blue dots) inside a transparent tracer particle (light blue circles) interferes with the unscattered portion to form a hologram that can be used to obtain the 3D locations of each marker through digital reconstruction. The velocity and vorticity of each tracer particle can be derived by tracking the motions of its internal markers.

### 2.1 Experimental setup based on digital inline holography

Digital inline holography (DIH) emerged more than two decades ago as a low-cost and compact technique for 3D microscopic imaging of biological samples (Xu et al. 2001). The particle tracking velocimetry (PTV) based on DIH, also known as DIH-PTV, has been applied for high-resolution 3D flow measurements, achieving significantly higher spatial resolution compared to the commercialized flow diagnostics techniques (e.g., V3V from TSI and tomographic PIV from LaVision) (Katz & Sheng 2010, Toloui & Hong 2015, Toloui, Mallery & Hong 2017). Particularly, it has been used to quantify the 3D rotation of a single nanorod (Cheong & Grier 2010), the orientation of a single ellipsoidal particle (Byeon et al. 2016), and the 2D rotation of a small number of irregular particles (Wu et al. 2015). In our proposed method, as illustrated in Fig. 1, the DIH setup consists of a 632-nm HeNe laser as the illumination light source and a high-speed CMOS camera (NAC Memrecam HX-5) with an objective imaging lens to capture the interference patterns (referred to as holograms hereafter) generated by the scattered light from each object within the sample volume and the non-scattered portion of the illumination laser.

The sample volume is filled with transparent fluid with suspended polymer microspheres with embedded particles for the proposed vorticity measurements. Such polymer microspheres have been broadly used in drug delivery (Oh et al. 2008), chemical sensing (Jiang et al. 2012), etc. Considering their clear optical property and stability for long-term use in liquid streams, they are used here as tracer particles, and they can be fabricated using a microfluidic approach to precisely control the size, composition, and internal structure. In our proposed method, the microsphere is made from polydimethylsiloxane (PDMS) using a flow-focusing microfluidic channel with a 70 µm x 70 µm square cross section (droplet generator chip fluidic 440, microfluidic ChipShop, Germany). An aqueous solution of sodium dodecyl sulfate (SDS, L37/71-Sigma Aldrich) in distilled water (2 wt.%) is used as the continuous phase. The dispersed phase is the liquid prepolymer PDMS (Dow Corning Sylgard 184 kit) with a base polymer to curing agent ratio of 6:4 mixed with silver coated glass particles (diameter 11 µm) serving as internal markers. The dispersed phase is injected into the main channel and then pinched by the continuous phase on both sides to form a droplet (Fig. 2a). Corona treatment is applied to the channels using a high-frequency generator (BD-20A, ElectroTechnic, IL) before use to ensure hydrophilic inner surfaces, which steadies the flow of the continuous phase (SDS solution) and ensures a > 90 contact angle between the PDMS prepolymer and the channel wall to facilitate the formation of PDMS droplets. The flow rates of the continuous and dispersed phases are adjusted independently using a delicate syringe pump (Chemyx Fusion 4000, Chemyx, TX) and are set at 0.032 mL/min and 0.0016 mL/min, respectively, to ensure droplets generated with the desired size in the jetting regime (Carneiro et al. 2016). The droplets are collected in a glass cuvette with the SDS solution and are thermally cured into solid tracers at 80°C for 30 min. The cured tracers are stored in the same SDS solution and are rinsed and redispersed into 60 wt.% glycerol aqueous solution for imaging under different flow conditions. The fabricated tracers are highly monodisperse with an average diameter of $100 \pm 2$ µm (Fig. 2b). Over 80% of the fabricated tracers have at least three spatially separated internal markers based on our microscopic imaging examination. To eliminate the optical distortion created at the tracer-fluid interface, the 60 wt.% glycerol aqueous solution is selected as the fluid medium for the current experiments, which yields a refractive index (RI) of 1.413, matching that of the tracers. Therefore, the internal markers (silver coated glass particles) within each tracer can be imaged non-distortedly for the holograms recorded at different distances from the focal plane (i.e., different $z$ planes) as illustrated in Fig. 2c. Moreover, the density of the tracers is $\rho_p = 0.97 \times 10^3$ kg/m$^3$, which is smaller than that of the fluid ($\rho_f = 1.15 \times 10^3$ kg/m$^3$). Such a configuration allows us to measure the vorticity change of small-scale vortices with the tracers potentially trapped at the center of rotation.

The traceability of our fabricated tracers is evaluated following the analysis described in Frish & Webb (1981). Specifically, the response time of the tracers undergoing translation ($\tau_u$) is calculated as $\tau_u = \frac{2}{9}r^2\rho/\mu \approx 0.54$ ms, where $r = 50$ µm is the radius of the tracers, $\rho_p = 0.97 \times 10^3$ kg/m$^3$ is the density of PDMS, and $\mu = 0.001$ kg/(m · s) is the viscosity of water at 20°C for generalizability. The response time of the tracers to the rotation ($\tau_\omega$) is estimated as $\tau_\omega = \frac{1}{15}r^2\rho/\mu$ based on the torque experienced by a rotating sphere in a fluid at a relative angular velocity $\Omega$ formulated by Chwang & Wu (1974). With our current tracer design, the vorticity response time is $\tau_\omega \approx 0.17$ ms. Considering the Stokes number $St = \tau_p/\tau_\eta < 0.1$ as a criterion for good traceability (Tropea, Yarin & Foss 2007), the fabricated tracers could follow the velocity and vorticity fluctuations in turbulent water flows generated in the laboratory with a Kolmogorov time scale, where $\tau_\eta > 5.4$ ms.

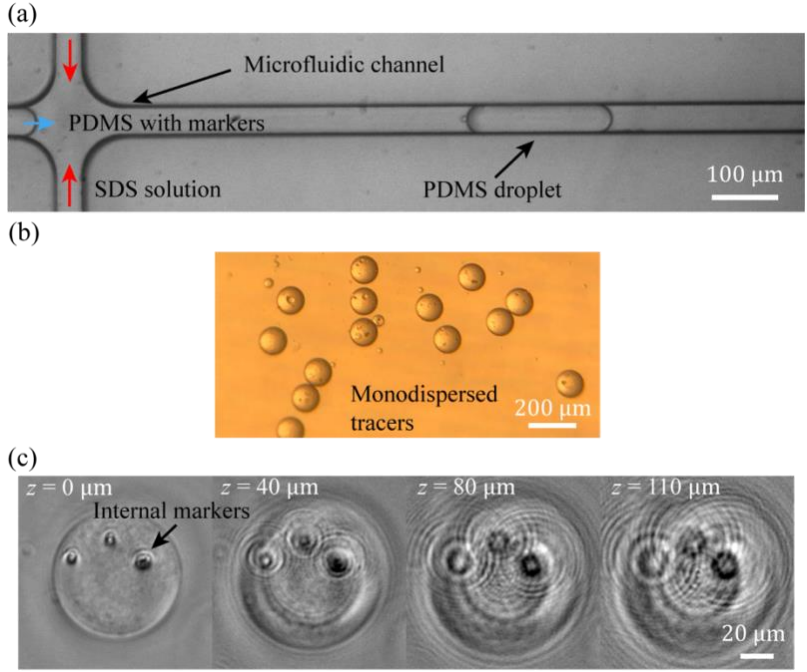

Figure 2: (a) PDMS droplet generation using a microfluidic channel, where the red arrows indicate the flow of the continuous phase (SDS solution) and the blue arrow indicates the flow of the dispersed phase (PDMS prepolymer mixed with markers). (b) The bright-field image of cured tracers with internal markers under the microscope with 4X magnification. (c) Holograms of one tracer with three internal markers at different $z$ depths where $z = 0$ μm represents the focal plane of the markers. The fringe becomes larger as the tracer moves further away from its focal plane.

## 2.2 Data processing method for vorticity measurements

The 3D locations of the internal markers for each tracer are obtained from the reconstructed 3D optical field from one recorded hologram using the regularized inverse holographic volume reconstruction (RIHVR) method by Mallery & Hong (2019). The RIHVR approach utilizes an inverse problem method for the volumetric reconstructions of particle fields as described by the equation (2.1):

$$\hat{x} = \underset{x}{\mathrm{argmin}}\{||Hx - b||_2^2 + \lambda R(x) \equiv f(x) + g(x)\}, \tag{2.1}$$

where $x$ is the optimal particle field to be found that minimized the difference between the captured hologram $b$ and the estimated hologram by transforming the particle field using an operation $H$, in which the Rayleigh-Sommerfeld diffraction kernel is used. The $\lambda R(x)$ term is the fused lasso regularization including a total variation regularization to enforce smoothness and a $l^1$ for sparsity of the reconstructed particle field. The algorithm is implemented with GPU processing which significantly reduces the computational cost. It has been successfully applied to the measurements of rotating rods in T-junction microchannels (Brady et al. 2009, Verrier et al. 2016, Endo et al. 2016, Jolivet et al. 2018), motions of microorganisms (You et al. 2020), and the flow in the viscous sublayer of wall-bounded turbulence (Kumar et al. 2021).

The reconstructed markers from a sequence of holograms are then linked to form 3D trajectories using open-source particle tracking software TrackPy (Allan et al. 2018). The markers in the same tracer are grouped together by implementing a simple k-means segmentation. The movement of the tracer in two consecutive images can be separated as the translation and the rotation. A 2D demonstration is shown in Fig. 3a. The translation ($d$) can be quantified as the spatial shift of the tracer centroid in the *x-y* plane (black arrow) between two consequent time frames $t_1$ and $t_2$. The rotation is the circular movement of internal markers around the $z$ axis, which is perpendicular to the *x-y* plane and passes through the center of the tracer. It can be quantified as the rotation angle $\theta$. Similarly, the tracer movement in a **3D** flow field can also be defined by the translation and rotation, except that the later now consists of three rotation angle components ($\theta_x$, $\theta_y$, $\theta_z$) in *x*, *y* and *z* directions, respectively. If a tracer starts at the location $L_1 = (p_{11}, p_{12}, p_{13}, \ldots)$ at $t_1$, where $p_{1i}$ is the 3D coordinate vector $(x_{1i}, y_{1i}, z_{1i})$ of the $i$ th marker, its location $L_2 = (p_{21}, p_{22}, p_{23}, \ldots)$ at $t_2$ can be predicted using the translation of the tracer centroid $d = (dx, dy, dz)$, and a guessed Euler angle $\theta = (\theta_x, \theta_y, \theta_z)$: $L_2' = L_1 + d + F(\theta, L_1)$, where $L_2'$ is the predicted marker location and $F(\theta, L_1)$ calculates the displacement of individual markers within the tracer from the rotation around the centroid by $\theta$. The rotation angles $\theta$ can be estimated by searching for the optimal rotation matrix **R** as functions of $\theta$ which minimizes the average end-point error of all markers ($\widehat{\mathbf{R}} = \underset{\mathbf{R}}{argmin} \frac{1}{N} \sum_{i=1}^{N} \epsilon_i^2$, where $N$ is the number of markers in the tracer, and $\epsilon_i$ is the end-point error for the $i$th marker, $\epsilon_i = |p_{2i}' - p_{2i}|$; or in the form of loss function in Wahba's problem (Wahba 1965), $\widehat{\mathbf{R}} = \underset{\mathbf{R}}{argmin} \frac{1}{2} \sum_{i=1}^{N} \alpha_i \| p_{2i}|_{O=\overline{L_2}} - \mathbf{R} p_{1i}|_{O=\overline{L_1}} \|^2$, where $\alpha_i$ is optional weights, $p_{ki}|_{O=\overline{L_k}}$ is the coordinates of the $i$th marker with the tracer center $L_k$ as origin). This problem is solved by using the singular value decomposition (SVD) method described by Markley & Mortari (2000). In either condition, the dimensionality of the optimization is reduced to three since we only solve for the rotation angles. The processing speed of the marker tracking, clustering, and rotation measurement steps is around 30 frames per second. The most computational-expensive part of our current method is the particle reconstruction using the RIHVR algorithm. It takes more than 2 min to finish the digital reconstruction of one 512 pixel × 512 pixel synthetic hologram. However, such speed is much faster comparing to the other compressive holography implementations (Brady et al. 2009, Verrier et al. 2016, Endo et al. 2016, Jolivet et al. 2018), yet slower than the tomographic particle reconstruction methods (Tan et al. 2019).

## 3. Results

### 3.1 Assessment with synthetic data

Our proposed algorithm was first assessed using synthesized holograms simulating different tracer rotation rates to determine optimal parameter settings for vorticity measurements and the corresponding uncertainty involved in the implementation of our algorithm. The synthetic holograms contain five tracers, each 100 μm in diameter with three to five internal markers (Fig. 3b & c). Assuming that the refractive index of the tracer material is closely matched with that of the surrounding medium, only the diffraction from the internal markers is simulated. We examine two rotation directions along two separate axes, i.e., rotation rate about the *y* axis ($\dot\theta_y$, *y*-rotation) and about the z axis ($\dot\theta_z$, *z*-rotation), with sample holograms presented in Fig. 3b & c, respectively. A total of 14 rotation rates varying from 0.05 to 1 rad/frame are simulated for both directions. A sequence of 100 holograms is synthesized for each rotation rate. The median of relative measurement errors ($\varepsilon_{\dot{\theta}}$) is shown in Fig. 3d and Fig. 3e for the *y*-rotation and *z*-rotation, respectively, with the error bar indicating the 50% confidence interval (CI). Relative measurement error is calculated as $\varepsilon_{\dot{\theta}} = |\dot{\theta} - \dot{\theta}_m|/\dot{\theta}$, where $\dot{\theta}$ is the rotation rate used for hologram synthesis and $\dot{\theta}_m$ is the measured Eulerian angle between each two consecutive frames.

In both cases, the errors first decrease as the rotation rate increases, reaching a minimum between 0.3 to 0.7 rad/frame, and then increase. This trend is caused by the errors introduced in location reconstruction and tracking when the rotation across two consecutive frames is either too low or too high, which hinders the accuracy of vorticity measurement. Specifically, for lower rotation rates, the error involved in determining the locations of markers within the tracer from 3D reconstruction is comparable to the small displacement of markers from the rotation across adjacent frames, which can lead to substantial uncertainty in the rotational rate measurements. Such an error reduces as the displacement grows with increasing rotation rate. However, when the marker displacement is as large as the physical distance between neighboring markers, the errors rise again due to the uncertainty involved in tracking the positions of each marker across frames. Moreover, the errors in the *y*-rotation case are, on average, larger than those in the *z*-rotation, with a minimal error of 1% for the former and 0.2% for the latter, respectively. This difference can be attributed to the fact that the *z* locations of the markers are less accurate compared to the *x* and *y* locations since the reconstructed 3D optical field from the hologram typically has a lower longitudinal resolution (*z* depth) by nature associated with the depth of field (DOF) (Katz & Sheng 2010).

The measurement errors exhibit similar trends for the synthesized holograms that simulate using microscopic objectives with two numerical apertures (NAs) (red circles for NA=0.15, and blue triangles for NA=0.25, respectively). These NA values are selected to match the ones used in the experiments presented in the following sections. In general, the errors with the larger NA are smaller on average. This trend is expected because the holograms with larger NA yield a more accurate longitudinal location (i.e., depth) after numerical reconstruction, which reduces the measurement errors associated with the tracking of markers.

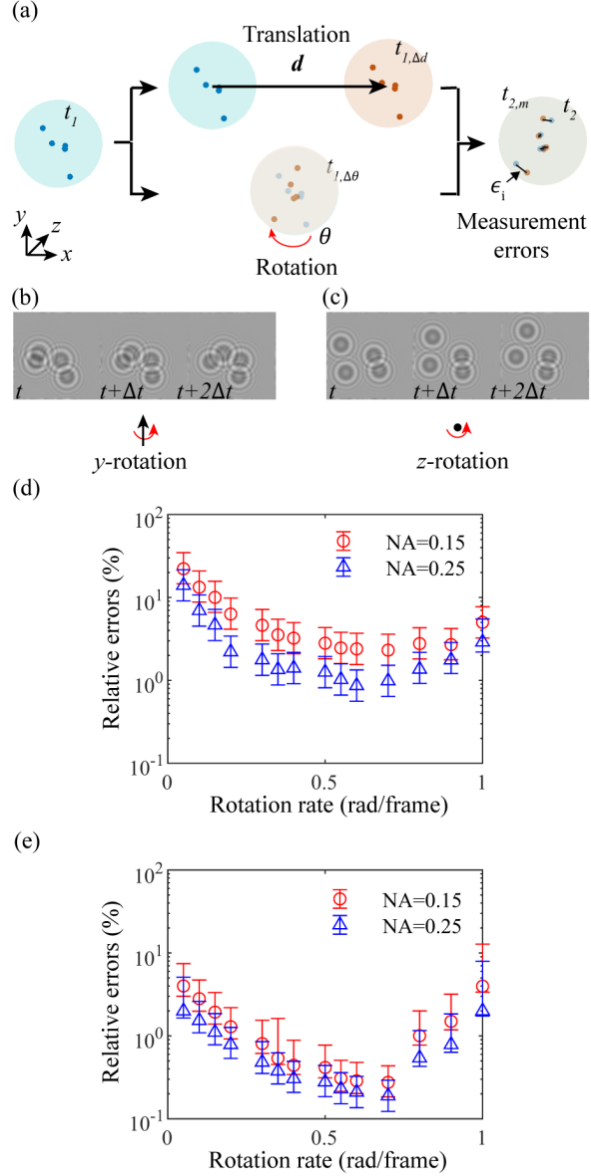

Figure 3: (a) A schematic illustration of the rotation measurement algorithm. The movement of a tracer at $t_1$ is divided into translation ($t_{1,\Delta d}$): the movement of the tracer centroid (solid black arrow), and rotation ($t_{1,\Delta\theta}$): the change of spatial locations of markers inside the tracer (solid red arrow). The error $\epsilon_i$ is defined as the distance between the predicted marker location ($t_{2,m}$) and the actual marker location at $t_2$. (b) Samples of three consecutive synthetic holograms to demonstrate the rotation about the $y$ axis and (c) the rotation about the $z$ axis with red arrows indicating the rotation direction. The median of the relative measurement errors as a function of rotation rates for (d) $y$-rotation and (e) $z$-rotations. The error bars represent the 50% confidence interval. The red circles represent errors with NA=0.15, and the blue triangles represent errors with NA=0.25.

In most experiment settings, it is likely that more tracers would be needed to reveal the flow structure. Thus, we generate additional synthetic data using 25 tracers (5 times of those in the previous synthetic cases) with 100-μm diameter in a 500 μm × 500 μm field of view to evaluate the algorithm. We set the NA to be 0.15 and rotation rate to be 0.3 considering the optimal measurement range and practical applications (with lower rotation rate). The sample synthetic holograms together with the reconstructed internal markers are presented in Fig. 4. A total of 100 synthetic images are generated and the same data processing steps are applied to generate the results. For the y-rotation case, the measurement uncertainty is 7±1%, and it is 2±0.3% for z-rotation case. The measurement uncertainties do increase comparing to the five tracer cases due to the larger errors in the tracking procedure with higher particle concentration.

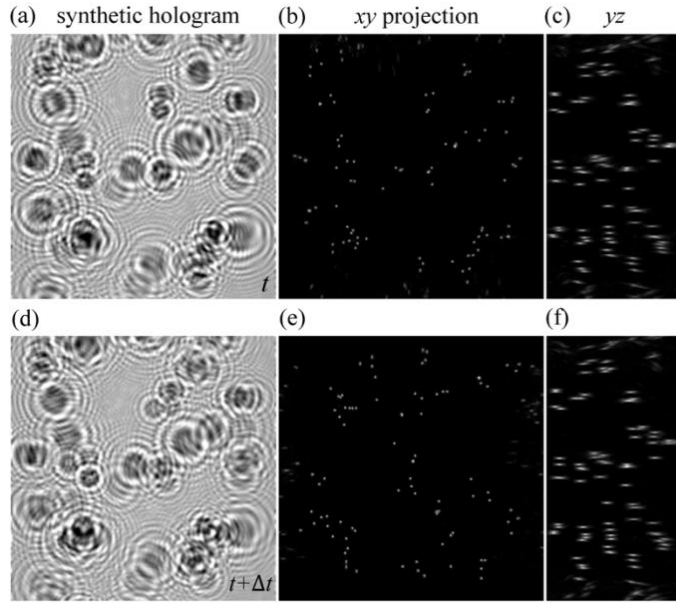

Figure 4: (a, d) Sample synthetic holograms with high tracer concentration at time $t$ and $t + \Delta t$ showing the rotation about the $y$ axis. The internal markers projected in the $xy$ plane (b, e) and $yz$ plane (c, f) from the reconstruction of the sample synthetic holograms.

**3.2 Assessment with a single tracer rotation**

A calibration experiment was conducted to assess the measurement uncertainty involved in the implementation of the proposed method in an actual refractive-index-matched (RIM) fluid medium. The holograms of a single tracer were acquired using the DIH setup shown in Fig. 5a. The tracer is glued to the center of a motorized spinning rod using a thin PDMS layer (Fig. 5b), which is submerged in the RIM glycerol aqueous solution. The spinning rod is 3D-printed using an SLA printer (Formlabs Form 3), and the diameter of the circular tip surface is 0.5 mm. This specific DIH setup (Fig. 5a) utilizes an additional spatial filter and a collimation lens with the HeNe laser to generate a coherent and collimated beam with a high-quality beam profile. Holograms (256 x 256 pixels) are captured by the high-speed camera at 1500 frames per second (FPS) after 20x magnification from an objective lens in a 0.7 mm × 0.7 mm field of view. The calibrated image resolution is 2.6 μm/pixel. Two consequent holograms are shown in Fig. 5c (left), where the fringe patterns of at least three markers are clearly visible within the tracer (light gray circle). The *x*-*y* projection of the reconstructed 3D location of each marker is presented on the right in Fig. 5c. The white dots represent the four identified markers that are used for tracking. The rotation rate of the motor is 900 revolutions per minute (RPM) (equivalent to 94 rad/s) to ensure the smallest displacement and longest trajectory. A total of 1200 holograms are captured, which are down-sampled for vorticity estimation at larger rotation rates. The relative measurement error is calculated as $\varepsilon_{\dot\theta} = |\dot\theta_{\text{motor}} - \dot\theta_m|/\dot\theta_{\text{motor}}$, where $\dot\theta_{\text{motor}}$ is the rotation rate of the motor and $\dot\theta_m$ is the measured rotation rate from two consecutive holograms.

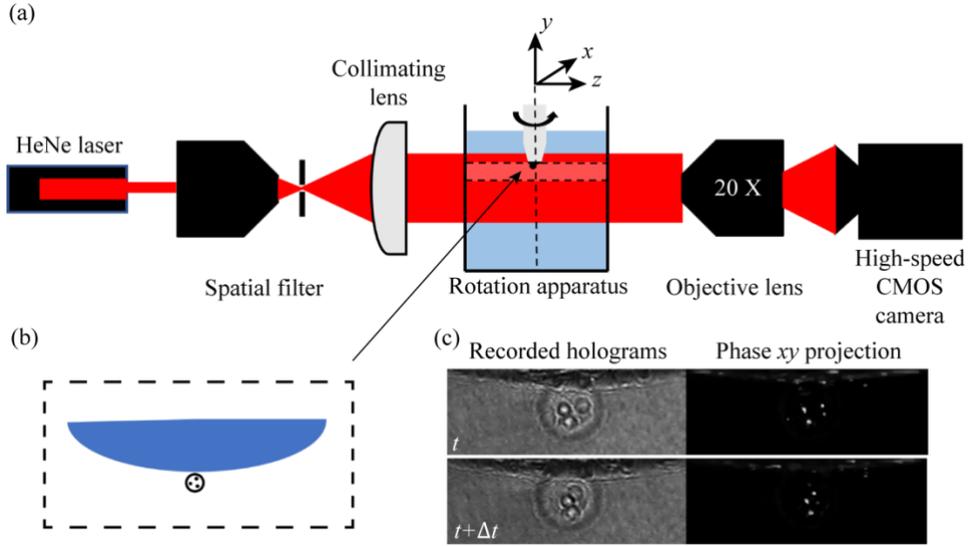

Figure 5: (a) Schematic of the DIH setup used in the calibration experiment. (b) An illustration of the position of the tracer (circle with dots inside) on the PDMS layer (blue area) attached to the 3D-printed rod. (c) Samples of the enhanced holograms (left) and the reconstructed *x-y* maximum intensity projection (right) of the markers inside the tracer.

Fig. 6 shows the median of the relative measurement errors for each rotation rate, with the error bars representing the 50% CI. As the rotation rate increase, the error first decreases and then increases with the minimum error at the rotation rate between 0.3 to 0.7 rad/frame. This trend is consistent with the results using the synthetic holograms, except that the errors are higher on average for the experimental holograms. Specifically, the minimal error in the experiments is about 2 % compared to that of 1 % for the synthetic holograms (Fig. 3d). Such larger errors are likely caused by the image distortion associated with the slight mismatch in the refractive indices between the medium and tracer particle (the light gray ring surrounding the internal markers), which lowers the accuracy of the marker location measurement. Moreover, compared to the synthetic holograms, the markers within the fabricated tracer can be less evenly distributed, and the noises in the experimental holograms are considerably higher due to the presence of dust particles in the sample volume and the aberration of the optical system. All these factors can cause larger uncertainties in the experiments.

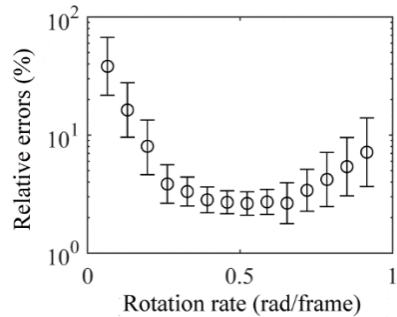

Figure 6: Median of the relative measurement errors at different rotation rates (black circles) for the calibration experiment with the error bars representing the 50% confidence interval.

### 3.3 Assessment with tracers in the flow under a rotating disk

We finally assessed the proposed method by measuring the vorticity of the flow generated by a rotating disk in a Lagrangian point of view. The DIH setup (Fig. 7a) is similar to that in the calibration experiment. The spinning rod attached to the motor is replaced by a 3D printed spinning disk with a 30-mm diameter. The disk is submerged in the aqueous glycerol solution and motorized to generate the von Kármán swirling flow in a transparent acrylic flow chamber with an internal size of 70 mm (length) x 70 mm (width) x 100 mm (depth). The solution inside is 80 mm deep. The kinematic viscosity of the glycerol solution is $\nu = 9.5 \times 10^{-6}$ m$^2$/s and the angular speed of the spinning disk is $\Omega = 198$ rad/s. Following the analysis in Section 2.1, the resulting response time of the tracer to the translation is estimated to be $\tau_u \approx 0.049$ ms, while the response time of the tracers to the rotation is $\tau_\omega \approx 0.015$ ms. Considering the time scale of the flow as $\tau_f = 1/\Omega = 5.1$ ms, the Stokes number of the tracers is calculated to be $St = \tau_u/\tau_f = 0.0096$, which satisfies the criterion for adequate traceability ($St \ll 0.1$, Tropea, Yarin & Foss 2007). The holograms (2560 x 1440 pixels) are captured under 3000 FPS to accommodate for the large flow speed. A total of 2,000 tracers are injected into the flow, and five are captured in the field of view (6.6 mm x 3.7 mm) with clearly visible markers inside. The same data processing procedure in the calibration experiment is used to calculate the Lagrangian velocity and vorticity along the trajectory of each tracer. The displacement and rotation rate of the tracers are measured every other frame to compensate for the high frame rate and ensure that the rotation rate is closer to the optimal range (i.e., 0.3-0.7 rad/frame).

The 3D trajectories of the five tracers are plotted as solid lines in Fig. 7b, with colors representing the velocity magnitude. The red arrows indicate the direction of the vorticity, and the arrow length represents the amplitude of the vorticities, which are twice the rotation rate. Both the velocity and vorticity increase as the tracers move closer to the rotation disk (the gray area on the top). To quantitatively evaluate the experimental measurements, the velocity and vorticity fields of the von Kármán swirling flow are simulated in MATLAB (details in Appendix A) for comparison. The vorticity in the plane parallel to the disk, $\omega_{\text{par,m}}$, is derived from two tracers with the longest trajectories and plotted as a function of the distance to the disk (triangle and circles in Fig. 7c). The error bars are the 50% confidence intervals at the measured vorticity. They are estimated through interpolation of the correlation function between the relative measurement error and the rotation rate (half of the vorticity) shown in Fig. 6. The theoretical values of the vorticity along the trajectories of the two tracers are calculated (dotted curves) for comparison using the equation, $\omega_{\text{par,th}} = \sqrt{\omega_r^2 + \omega_\theta^2}$, where $\omega_r$ and $\omega_\theta$ are derived as $\omega_r = -r\sqrt{\Omega^3/\nu}G'$ and $\omega_\theta = r\sqrt{\Omega^3/\nu}F'$, and $r$ is the radial distance of the tracers to the disk center, $G'$ and $F'$ are the first-order derivatives of the self-similar variables which are functions of the distance to the disk, $y$ (see Appendix A). Due to the large velocity relative to vorticity in the flow, the maximum rotation rate is around 0.15 rad/frame with a displacement of around 50 pixels. Even though the rotation rate is not in the optimal range (0.3 – 0.7 rad/frame), the experimental measurements fit well with the analytical solutions exhibiting a growing rotation rate as the tracers approach the disk. The theoretical values are all within the 50% confidence intervals (error bars) of the measured rotation rates along the two trajectories. This measurement can be further improved by running the disk at a higher angular speed to ensure a higher rotation rate within the optimal range.

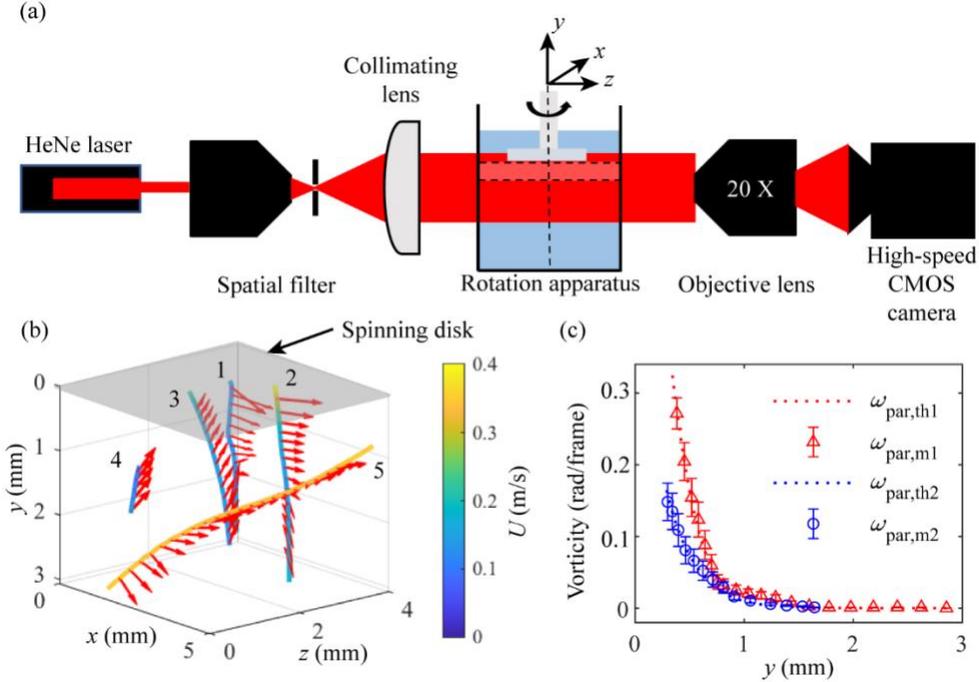

Figure 7: (a) Schematic showing the setup for the vorticity measurements of von Kármán swirling flow. (b) Side view of the 3D trajectories of five captured tracers. The color of the solid line represents velocity magnitude, and red arrows correspond to the measured vorticity vectors. (c) The vorticity of two tracers at different distances ($y$) from the rotating disk. The theoretical results were plotted as the dotted lines, and the experimental measurements were plotted as symbols (triangles as tracer 1 and circles as tracer 2). Error bars represent the 50% confidence interval of the measurements.

## 4. Conclusions and discussion

We develop a new method for direct vorticity measurement in fluid flows based on digital inline holography (DIH). The DIH system utilizes a collimated laser beam to illuminate the tracers with internal markers and a digital sensor to record the generated holograms. The tracers made of the polydimethylsiloxane (PDMS) prepolymer mixed with internal markers are fabricated using a standard microfluidic droplet generator. A rotation measurement algorithm is developed based on the 3D location reconstruction and tracking of the internal markers and is assessed through synthetic holograms to identify the optimal parameter settings and measurement range (e.g., rotation rate from 0.3 to 0.7 rad/frame under numerical aperture of imaging of 0.25). Our proposed method based on DIH is evaluated by a calibration experiment of single tracer rotation, which yields the same optimal measurement range. Using von Kármán swirling flow setup, we further demonstrate the capability of the approach to simultaneously measure the Lagrangian rotation and translation of multiple tracers.

Compared with other state-of-the-art techniques for 3D flow measurement, our method can achieve a better resolution for the vorticity measurement as it requires no spatial derivative and interpolation. The density of the tracers should be smaller than that of the surrounding fluid so that they can potentially be trapped in the core of vortices for vorticity measurements in a highly localized region on the scale less than 100 μm. Our current approach has demonstrated the measurements using $100-$ μm tracers. The microfluidic droplet generator can be adjusted to yield a tracer size as small as tens of microns by adjusting the channel size and flow rates of the continuous and dispersed phases (Carneiro et al. 2016), which will lead to further improvements in the measurement resolution. The precise measurement of vortices in a localized microscale region is important for understanding the spatial organization of the Kolmogorov scale vortex filaments (although the width of these tube-like vortex filaments are on the order of Kolmogorov scale, their length can be much larger as shown in She et al. 1990) and their dynamics which influences the spatial and temporal characteristics of turbulence dissipation. Moreover, in comparison to the existing rotation measurement approach, our measurement system can quantify the rotation of a large number of tracers simultaneously as demonstrated in our high concentration synthetic tracer tests (100,000 tracers/$cm^3$, considering the limitation of the marker cluster segmentation). This concentration level can be further improved by employing smaller tracers, giving us the opportunity to probe into the geometric structures of turbulence and their temporal characteristics at different scales as well as the interaction among these structures in the flows. In addition, compared with standard laser-based flow measurement systems, except for the high-speed camera, our DIH setup is low-cost and compact, requiring only a low-power laser. Using a standard microfluidic droplet generator, our tracer preparation procedures are relatively straightforward and stable compared with the method described in Wu et al. (2018). It is more generalizable to fabricate tracers with different sizes and anisotropic internal marker configurations.

However, there are also certain limitations to the proposed method. First, the rotation measurement algorithm involves several steps (i.e., 3D reconstruction, tracking, marker clustering, rotation matching, etc.), and the uncertainties accumulate from each step. This limitation can be addressed by developing a deep-learning-based image registration method that derives the displacement and rotation directly from the correlation of adjacent frames, similar to particle image velocimetry. Deep learning, especially convolutional neural networks (CNNs), can be a powerful tool for such image correlation and has been applied for optical flow and PIV measurements (Ilg et al. 2017, Cai et al. 2019, Teed & Deng 2020). Future work can focus on the feasibility of CNNs for direct measurement of velocity and vorticity from a sequence of holograms. With a specific design in the feature extraction layers of the CNN, the effects of image noise can be reduced, and the measurement would be more robust for the experiment images. Second, in the demonstration experiment, the system captures a limited number of tracers in the field of view, hindered by our current seeding procedure. The low tracer concentration limits how well our proposed approach can resolve the spatial variation of vorticity in space (i.e., the spacing between neighboring vorticity vectors). By improving the control of the tracer seeding step, we can increase the concentration of tracers in the sample volume to ensure the spatial resolution needed for the vorticity field measurements. Last, since our method is based on microscopic imaging, its application in large-scale measurements (centimeter-scale or larger) is limited. However, it is possible to integrate our proposed DIH measurement with other state-of-the-art 3D PTV techniques such as tomographic, defocusing, and plenoptic imaging for simultaneous measurements both on small and large scales due to the compact optical setup.

**Appendix A**

The von Kármán swirling flow can be described by the equations below (Equations A.1 to A.4), assuming steady flow as the image data is taken long enough after turning on the motor:

$$\frac{2u}{r} + \frac{dw}{dz} = 0, \qquad (A.1)$$

$$\left(\frac{u}{r}\right)^2 - \left(\frac{v}{r}\right)^2 + w\frac{d(u/r)}{dz} = \frac{-1}{\rho}\frac{\partial p}{\partial r} + \nu\frac{d^2(u/r)}{dz^2}, \tag{A.2}$$

$$\frac{2uv}{r^2} + w\frac{d(v/r)}{dz} = \nu\frac{d^2(v/r)}{dz^2}, \tag{A.3}$$

$$w\frac{dw}{dz} = \frac{-1}{\rho}\frac{\partial p}{\partial z} + \nu\frac{d^2 w}{dz^2}, \tag{A.4}$$

with the boundary conditions for fluid with $z > 0$ are:

$$u = 0, v = \Omega r, w = 0, p = p_0 \text{ for } z = 0,$$

$$\text{and } u = 0, v = 0 \text{ for } z \to \infty.$$

In these equations, $(u, v, w)$ are the velocity components in cylindrical coordinates $(r, \theta, z)$, with the origin at the center of the spinning disk and $(r, \theta)$ representing the plane parallel to the disk. Note that the coordinate system is defined in a way different from that of the image data. The cross section of the measurement sample volume is around 6.6 mm x 3.7 mm, much smaller compared to the size of the flow chamber, and it is only around one fifth of the diameter of the disk. With the small sample volume, it is safe to assume that the boundaries of the flow chamber have little influence on the flow in the measurement volume and there is no rotation at infinity. Thus, the pressure becomes independent of $r$. By introducing the transformation: $\eta = \sqrt{\frac{\Omega}{\nu}}z$, $u = r\Omega F(\eta)$, $v = r\Omega G(\eta)$, $w = \sqrt{\nu\Omega}H(\eta)$, $p = p_0 + \rho\nu\Omega P(\eta)$, where $\nu$ is the kinematic viscosity of the fluid and $\Omega$ is the angular speed of the spinning disk, the self-similar equations (Equation A.5 to A.8) can be derived:

$$2F + H' = 0, \tag{A.5}$$

$$F^2 - G^2 + F'H = F'', \tag{A.6}$$

$$2FG + G'H = G'', \tag{A.7}$$

$$P' + HH' - H'' = 0, \tag{A.8}$$

and the boundary conditions for fluid with $\eta > 0$ are now:

$$F = 0, G = 1, H = 0, P = 0 \text{ for } \eta = 0,$$

$$\text{and } F = 0, G = 0 \text{ for } \eta \to \infty.$$

Moreover, the vorticity components in the cylindrical coordinates can also be derived as follows:

$$\omega_r = -r\sqrt{\Omega^3/\nu}G', \tag{A.9}$$

$$\omega_\theta = r\sqrt{\Omega^3/\nu}F', \tag{A.10}$$

$$\omega_z = 2\Omega G. \tag{A.11}$$

The self-similar equations (Equations A.5-A.7) were numerically solved using the bvp4c function in MATLAB to obtain the profiles of functions $F$, $G$, and $H$ as well as the derivatives $F'$ and $G'$. By applying $\nu = 9.5 \times 10^{-6}\,\text{m}^2/\text{s}$ and $\Omega = 198\,\text{rad/s}$ to the equations for $(u, v, w)$ and $(\omega_r, \omega_\theta, \omega_z)$, the analytical velocity and vorticity fields can be derived. To compare with the measurement, the vorticity is converted to rotation rate by dividing the imaging frame rate and two as the vorticity is twice the angular velocity.

**Acknowledgements**

This study is supported by the Army Research Office (Program Manager, Dr. Matthew Munson) under the award No.W911NF2010098. The authors would like to thank Keven Mallery, Tong Zhou, and Rafael Grazzini Placucci for their help and support during the course of this research.